\documentclass[showpacs,preprintnumbers,superscriptaddress,amsmath,reprint,aps,pra,twocolumn]{revtex4-1}

\usepackage{bm}
\usepackage{graphicx} 
\usepackage{color}
\usepackage{amsmath}
\usepackage{booktabs}

\newcommand{\EEA}{\end{eqnarray}}
\newcommand{\EEAN}{\end{eqnarray*}}

\begin{document}
 
\title{Universal temperature dependence of the ion-neutral-neutral three-body recombination rate}

\author{Jes\'{u}s P\'{e}rez-R\'{i}os}

\affiliation{School of Natural Sciences and Technology, Universidad del Turabo, Gurabo, PR00778, USA }

\author{Chris H. Greene}

\affiliation{Department of Physics and Astronomy, Purdue Unviversity, West Lafayette, Indiana 47907  USA }

\date{\today}

\begin{abstract}

A classical approach based on hyperspherical coordinates is used to derive a first principle formulation of the the ion-neutral-neutral three-body recombination rate, A$^+$ + A + A $\rightarrow$ A$^+_2$ + A, in terms of the mass of the atom and its polarizability. The robustness and predictive power of our approach have been checked in comparison with experimental data of rare gas three-body recombination as well as previous theoretical frameworks, which need one or two atom-dependent fitting parameters. Thus, our approach is general and applicable to any ion-atom-atom system.

\end{abstract}

\maketitle

\section{Introduction}

Three-body recombination or three-body association is a chemical reaction in which a molecule emerges as the product state after a three-body encounter, i.e., A + A + A $\rightarrow$ A$_2$ + A. This reaction is present in different disciplines of physics and chemistry, for instance, in astrophysics,  hydrogen three-body reaction plays a major role on star formation due the cooling properties of H$_2$ owing its internal degrees of freedom~\cite{Flower2007,Bovino2014}, or in ultracold physics, being three-body recombination one of the main atom loss mechanism in Bose-Einstein condensates~\cite{Hess1983,Hess1984,Goey1986,Fedichev1996,Burt1997,Esry1999}.  

\begin{figure}[h]
\centering\includegraphics[width=0.7\columnwidth]{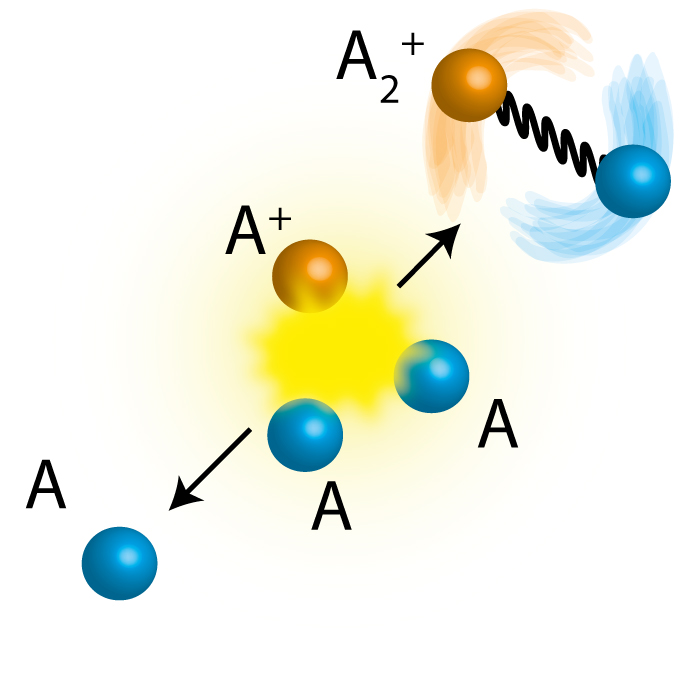}
\caption{Cartoon representation of ion-neutral-neutral three-body recombination. }
\label{fig1}
\end{figure}

When one of the three colliding partners is an ion, three-body recombination leads to the formation of a molecular ion most of the times, i.e. A$^+$ + A + A $\rightarrow$ A$^+_2$ + A, see Fig.\ref{fig1} for a cartoon representation of the three-body recombination. This reaction when $A$ is a rare gas is of fundamental interest in radiation physics, concretely in gaseous radiation detectors~\cite{Neves2007,Neves2010}, excimer lasers~\cite{Jones1980,Papanyan1995} and spectrometers~\cite{Eiceman2005}. Similar reactions involving alkali atoms play an important role in cold chemistry~\cite{Krukow2016}, where the product of the reaction is a weakly bound molecular ion that ulteriorly relaxes due to the collisions with the neutral atoms~\cite{JPR2018}.

Ion-neutral-neutral three-body recombination has been studied from several theoretical frameworks. One of the earliest treatments of this reactions employed the detailed balance condition in dissociation processes to obtain the corresponding association rate, leading to a the three-body recombination rate $k_3 \propto T^{-1}$~\cite{Niles1965,Niles1965bis}, explaining qualitatively some of the experimental data at that time. Later on, this reaction was approached from  an 'indirect' approach: in which a three-body process is viewed as a two step mechanism~\cite{Mahan1965,Smirnov1966,Dickinson1971}. The first step is a two-body event leading to the formation of a resonant complex, which eventually will be stabilized in the second step through a collision with a third body. Different temperature dependence of the rate may be obtained by means of this approach since it strongly depends on the way the resonant complex is described and the stabilization probability. In particular, when a capture model is employed for the resonant complex formation, $k_\text{3} \propto T^{-3/4}$, which turns out to be more accurate in comparison with the available experimental data. On the contrary, if the population of the resonant complex is described by assuming thermal equilibrium, one finds a more intricate relationship between $k_\text{3} $ and $T$, although more accurate from the qualitative and quantitative perspective in comparison with the experimental data. In the same vein, some quantal calculations following the same logic have revealed a great accuracy in describing He$^+$-He-He recombination~\cite{Xie2003}.

Most of these theories for ion-neutral-neutral three-body recombination depend on some 'free' parameters that need to be fitted for each atomic specie in order to reach a proper description of the underlying physics. Thus, a more general treatment of this fundamental chemical process is needed. Recently, we have developed a 'direct' three-body approach based on a Newtonian approach of the dynamics by means of hyperspherical coordinates~\cite{JPR2015}, leading to $k_3\propto T^{-3/4}$, which has been experimentally corroborated at cold temperatures $T\lesssim 1$K~\cite{Krukow2016}, as well as numerically. This fuels us to go one step beyond and generalize our approach to derive an analytical an general expression for the ion-neutral-neutral three-body recombination rate, depending only on intrinsic properties of the colliding atoms.

In this paper, we present a ``direct'' three-body approach based on a previously derived hybrid hyperspherical-classical trajectory calculations method, which naturally leads to a realistic description of the experimental data for ion-neutral-neutral three-body recombination of rare gases. The derived three-body recombination rate only depends on intrinsic properties of the rare gas atoms: mass and polarizability, thus being a general and parameter free approach to ion-neutral-neutral three-body recombination.

\begin{figure*}[t]
\centering\includegraphics[width=2.0\columnwidth]{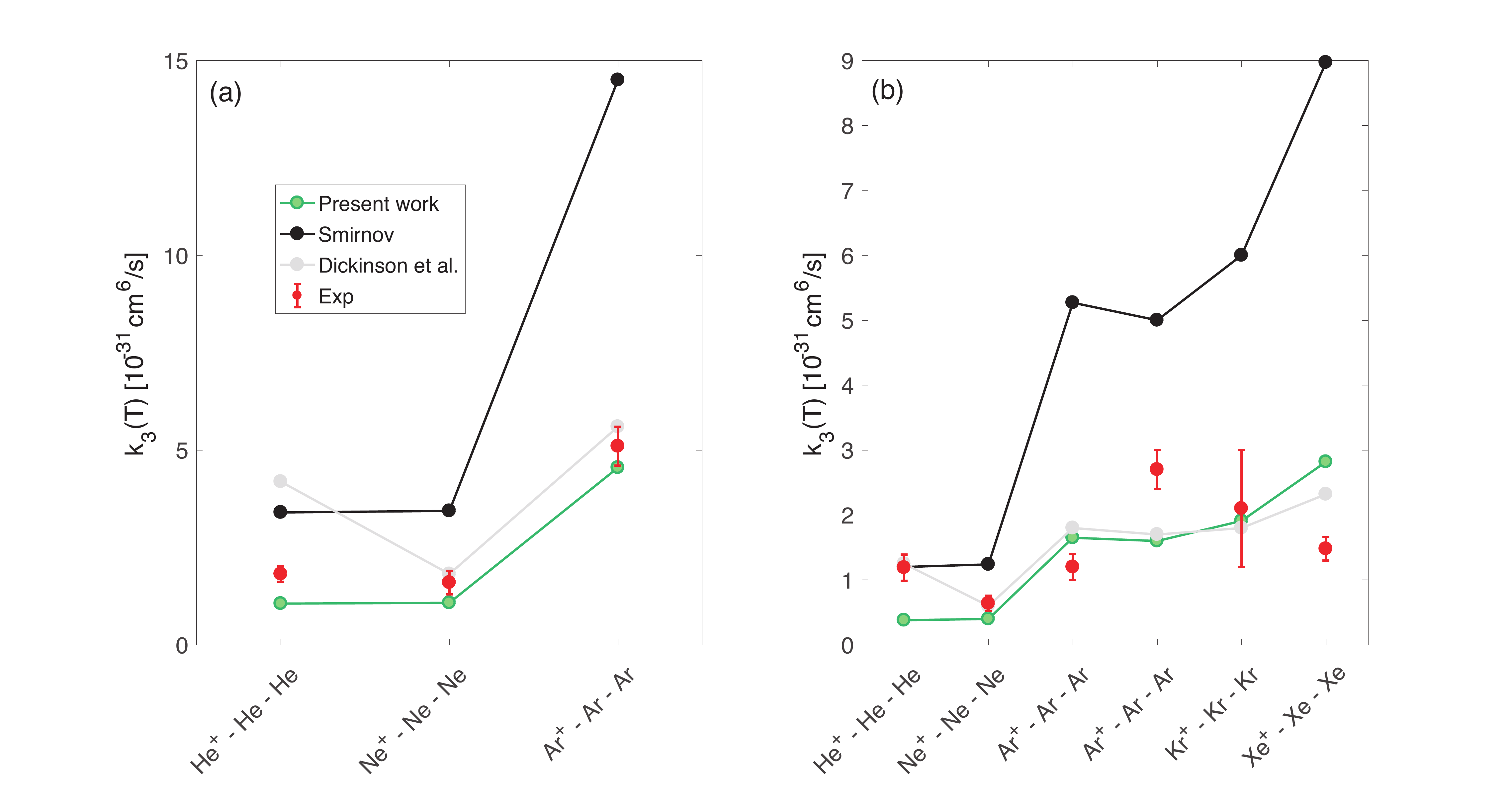}
\caption{Ion-neutral-neutral three body recombination rate for different rare gas atoms and different temperatures. In panel (a) the temperature is 78 K. In panel (b) the temperature for all the rare gas is 300 K except the second Ar$^+$-Ar-Ar data point, which is for 320K. The experimental data points of panel (a) have been obtained from Ref.~\cite{Johnsen1980}, for panel (b) the data points are taken  from Ref.\cite{Johnsen1980} for He, Ne and Ar, whereas Kr and Xe data are taken from Ref.\cite{Neves2010}. In panel (b) there are two data points for Ar, the first one (from left to right) corresponds to 300~K~\cite{Neves2010}and the second to 320 K~\cite{Johnsen1980}. }
\label{fig2}
\end{figure*}

\section{Model}

Here we adopt the previously developed classical description of three-body collision processes using hyperspherical coordinates~\cite{JPR2014,RMP}, within this theoretical framework, the three-body recombination cross section is given by~\cite{JPR2014}

\begin{equation}
\label{eq1}
\sigma_{\text{3}}(E_\text{k})=\frac{8 \pi^2}{3}\int_{0}^{b_\text{max}(E_{\text{k}})}b^4db,
\end{equation}

\noindent
where $b_\text{max}(E_{\text{k}})$ represents the maximum impact parameter for the three-body recombination reaction, which depends on the kinetic energy $E_{\text{k}}$. In this equation it is assumed that every trajectory whose impact factor is equal or smaller than $b_\text{max}(E_{\text{k}})$ will lead to a three-body recombination event, analogous to a Langevin-type hypothesis. Therefore, Eq.~(\ref{eq1}) represents the maximum cross section for a given kinetic energy.

For ion-atom-atom three-body recombination processes the ion-atom interaction mainly dictates the fate of three-body events~\cite{JPR2015,Krukow2016}, since the ion-atom interaction is longer-ranged than the usual van der Waals forces between atoms. In the framework of the Langevin capture model for ion-atom collisions, an inelastic collision or chemical reaction will happen when the kinetic energy, $E_\text{k}$, is larger than the height of the potential barrier, which determines the so-called Langevin impact parameter, $b_{L}=(2\alpha/E_k)^{1/4}$, where $\alpha$ stands for the atom polarizability. This represents the maximum impact parameter for a given inelastic ion-atom collision, whereby Eq.(\ref{eq1}) reads as

\begin{equation}
\label{eq2}
\sigma_{\text{3}}(E_\text{k})=\frac{8 \pi^2}{3}\int_{0}^{b_L}b^4db,
\end{equation}

\noindent
and performing the integration one gets

\begin{equation}
\label{eq3}
\sigma_{\text{3}}(E_k)=\frac{8 \pi^2}{15}\left(\frac{2\alpha}{E_k} \right)^{5/4}.
\end{equation}

The energy-dependent three-body recombination rate is defined as

\begin{equation}
k_{3}(E_k)= \sigma_{3}(E_k)\left( \frac{2E_k}{\mu}\right)^{1/2},  
\end{equation}

\noindent
and performing its average through the Maxwell-Boltzmann distribution one finds

\begin{eqnarray}
k_{3}(T)&=&\frac{1}{2(k_BT)^3}\int_{0}^{\infty} k_{3}(E_k)E_k^{2} e^{-\frac{E_k}{k_BT}}dE_k \nonumber \\
&=&\frac{\Gamma(9/4)3^{1/4}}{\sqrt{2}(k_BT)^{3/4}}\frac{\left( 2\alpha \right)^{5/4}}{\sqrt{m}}.
\end{eqnarray}

\noindent
As expected, we find that the three-body recombination rate depends on the temperature as $T^{-3/4}$~\cite{JPR2015,RMP}. It is worth noticing that  the same dependence on the temperature was obtained by Smirnov and~\cite{Smirnov1966} back in the 60's but assuming that the three-body recombination can be described as two different two-body collisions. The first leads to the formation of a resonant complex and the second may stabilize this complex to the formation of the molecular ion. 

\section{Results and discussion}

The theoretical approach presented in the previous section has been applied to ion-neutral-neutral three-body recombination of rare gas atoms, and its predictions compared with experimental data in Fig.\ref{fig1}, where panel (a) presents results for data at 78K whereas panel (b) shows data for 300 K. The present theory agrees fairly well with the room temperature data, but it does extremely well at low temperatures. Moreover, our approach describes qualitatively the dependence of the rate on the properties of the atom at hand, independently of the temperature.

In Fig.~\ref{fig2}, two more theoretical results based on the ``indirect'' approach are shown as well. One of them is due to Smirnov~\cite{Smirnov1966} which employs a capture model for the first two-body encounter wighted by the ratio between stabilization to dissociation collisions of the intermediate complex. This approach leads to a $k_{3}(T)\propto T^{-3/4}$, as in our derivation, but in this case there is a free parameter that needs to be fitted based on the atomic specie at hand. In particular, we have chosen the same value of this parameter as in Smirnov's original work~\cite{Smirnov1966} which was specially calculated for He$^+$ - He - He at room temperature, and it shown as the black solid line in Fig.\ref{fig2}. Smirnov's approach gives an excellent agreement in comparison with the experimental data for He, as panel (b) of Fig.\ref{fig2} shows, however its predictive level for the rest of the rare gases is just qualitative, independently of the temperature.

The approach of Dickinson et al.~\cite{Dickinson1971} (referred to as Dickinson's approach for brevity) is more involved than Smirnov's version since the authors consider explicitly the stabilization probability as a function of the internal state of the intermediate complex. Within this approach the three-body recombination rate is given by

\begin{eqnarray}
\label{Dickinson}
k_{3}(T)=\frac{\pi^2\hbar\alpha}{\sqrt{\mu m}k_BT}\bigg \{\log{\left(\frac{m^2\alpha k_B T}{\hbar^4}\right)} -\gamma \nonumber \\
+4\rho\Gamma(1/4)\left(\frac{2k_BT}{\alpha}\right)^{1/4}+ 4\rho^2\sqrt{\frac{2\pi k_B T}{\alpha}} \nonumber \\
 -2\rho^2(J_M+1/2)^2\frac{2\hbar^2}{m\alpha}- \frac{2\rho\hbar}{\sqrt{m \alpha}} (J_M+1/2) \nonumber \\
 +\frac{\hbar^4}{8m^2\alpha k_B T}(J_M+1/2)^4\bigg \}
\end{eqnarray}

\noindent
where $\gamma$ is the Euler's constant, $\rho $ is the averaged distance of closest approach of the third body and $J_M$ stands for the maximum angular momentum state in which on average there is quasi-bound state for a $1/r^4$ long-range interaction~\cite{Dickinson1971}. Thus, $\rho$ and $J_M$ are parameters that depends on the system at hand as well as its temperature. However, the value of $\rho$ is taken arbitrarily to be 6~a$_0$ independently of the gas at hand and $J_M\sim 10$. The results of Dickinson et al. are shown as the grey line of Fig.~\ref{fig2}, and they describe qualitatively all the experimental data, and even quantitively the room temperature data [panel (b)]. Dickinson's approach appears to be comparably successful in describing the experimental data as the present approach.

\begin{figure}[t]
\centering\includegraphics[width=1.0\columnwidth]{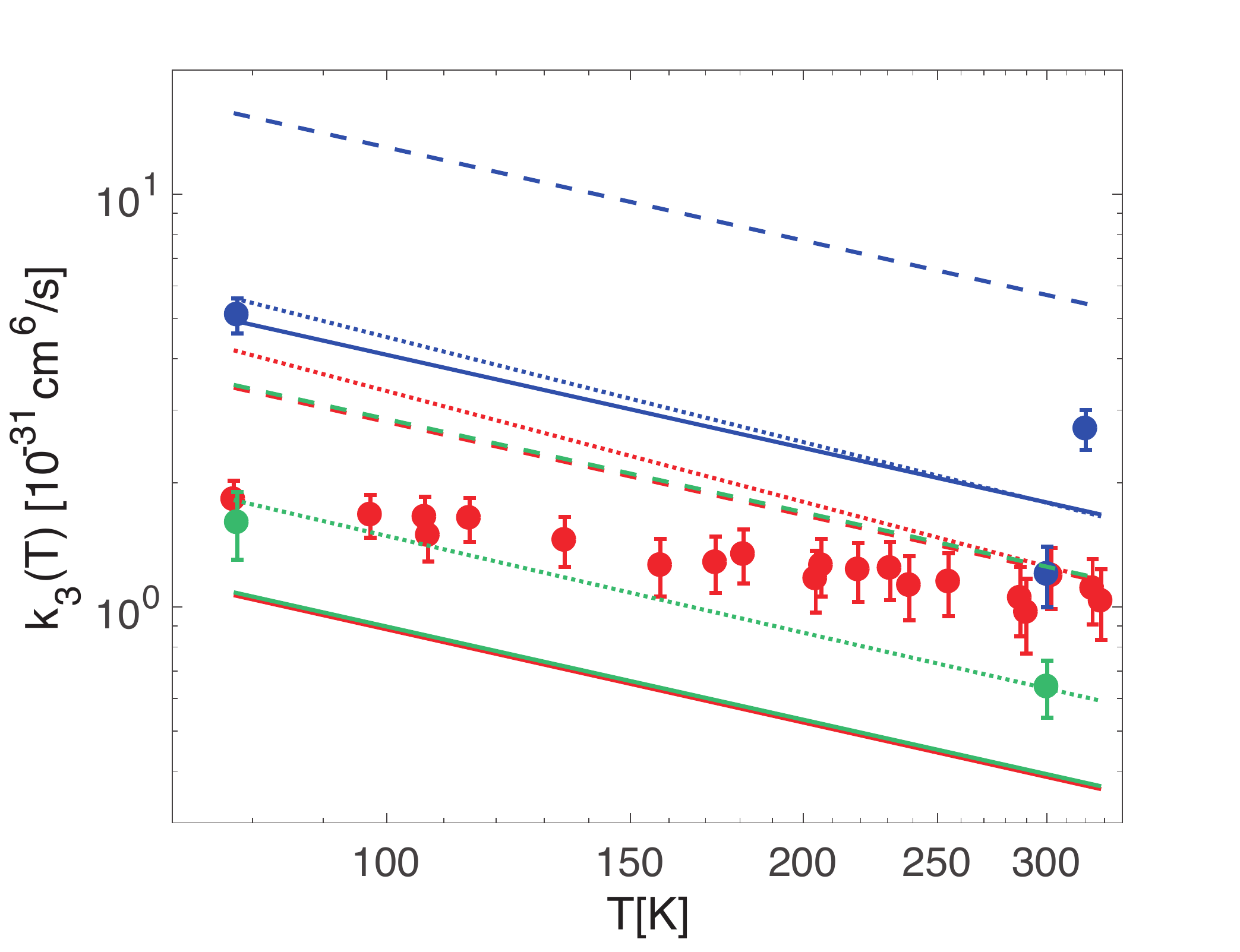}
\caption{Temperature dependence of the ion-neutral-neutral three-body recombination of rare gases. The experimental data are represented by the solid symbols, in red for He, green for Ne and blue for Ar. The solid lines following the same color coding stand for the results of our model, whereas the dashed line is for the Smirnov's approach and the dotted for the model of Dickinson et al. (see text for details).The experimental data  have been obtained from Ref.\cite{Johnsen1980} with the exception of the Ar data point at 300 K which is taken from Ref.\cite{Neves2010}.}
\label{fig3}
\end{figure}

However, a closer look into Eq.~(\ref{Dickinson}) shows terms $\propto T^{-2}$ and $\propto T^{-1}$  that ultimately will lead to extremely large rates at low temperatures, as it is noticed in Fig.~\ref{fig2}. Moreover, recent cold chemistry experiments seem to rule out such a temperature dependence~\cite{Krukow2016}. To study this further, Fig.~\ref{fig3} shows the ion-neutral-neutral three-body recombination rate as a function of the temperature for He, Ne and Ar. Dickinson's approach is represented by the dotted line, the dashed line stands for Smirnov's model and the solid line is our present approach. From Fig.~\ref{fig3} one observes that the present approach describes properly the temperature dependence of the rate although the quantitative agreement is only reached for Ar. Dickinson's approach shows an overall excellent agreement with the experimental data, however for the 78 K data it exhibits an incipient deviation to larger rates which is pathological to the approach. Finally, Smirnov's approach describes extremely well the He data for T~$\gtrsim 250$~K, but for the rest of the cases it fails. 

When plotting the different experimental data for Ar$^+$ + Ar + Ar $\rightarrow$ Ar$^+_2$ + Ar we have noticed that the rate for 300 K~\cite{Neves2010} lies below the rate at 320 K~\cite{Johnsen1980} which is hard to explain, since the three-body recombination decreases as the collision energy increases. Another possibility, could be the presence of resonances, however at room temperature collision many partial waves contribute to the scattering and hence any resonance effect will normally be washed out. Therefore, it seems that some extra work needs to be invested in the Ar three-body recombination problem at T$\gtrsim$ 300 K to solve this apparent inconsistency.

\section{Conclusions}

A recent direct classical-based approach for three-body recombination has been applied to ion-neutral-neutral recombination of rare gas atoms, and as a result an analytic and universal expression for the three-body rate has been obtained. Our results have been checked against the available experimental data, as well as different available theoretical approaches, confirming the accuracy of our approach and showing that our model captures the most relevant physics behind three-body recombination.

Some other models for ion-neutral-neutral three-body recombination have been developed, although they assume that two distinct two-body processes act the same as a three-body event, but those need the inclusion of one or two different fitting parameters that depends on the atom at hand. These theories can describe the general behavior of the three-body recombination rate, however they fail to describe the three-body physics at low temperatures. Therefore, our approach seems to be more general and robust than previous ones.

Finally, our analysis has allowed us to perceive some troubling discrepancies in the experimental data of Ar$^+$ + Ar + Ar $\rightarrow$ Ar$^+_2$ + Ar measured by two different groups. We hope this will help to motivate experiments on this reaction to clarify what is the proper behavior of the three-body recombination rate for ion-neutral-neutral collisions. 

\section{Acknowledgements}

J. P.-R. acknowledge the hospitality of KITP during part of the preparation of this work. The work of CHG has been supported by the U.S. Department of Energy, Office of Science, grant number DE-SC0010545.
\bibliography{3B}

\end{document}